\documentclass[pdftex,aip,apl,reprint]{revtex4-1}
\usepackage{amsmath}
\usepackage{graphics}
\usepackage[pdftex]{graphicx}
\usepackage{wasysym}
\usepackage{setspace}
\usepackage{color}
\usepackage{nicefrac}
\usepackage{braket}
\usepackage{amsfonts}
\usepackage{amssymb}
\newcommand{\figref}[1]{Fig.~\ref{fig:#1}}
\newcommand{\Figref}[1]{Figure~\ref{fig:#1}}

\renewcommand{\eqref}[1]{Eq.~(\ref{eq:#1})}

\newcommand{\citeasnoun}[1]{Ref.~\onlinecite{#1}}

\def\a{s}
\def\b{s}
\newcommand{\add}[1]{\if\a\b{{\color{red} #1}}\else{#1}\fi}
\newcommand{\comm}[1]{\if\a\b{{\color{blue}\{\small \sc #1\}}}\else{}\fi}
\newcommand{\del}[1]{{\if\a\b{{\color{magenta}[[#1]]}}\else{}\fi}}

\begin{document}

\title{Designing evanescent optical interactions to control the
  expression of Casimir forces in optomechanical structures}

\author{Alejandro W. Rodriguez}
\affiliation{School of Engineering and Applied Sciences,
Harvard University, Cambridge, MA 02139}
\affiliation{Department of Mathematics, 
Massachusetts Institute of Technology, Cambridge, MA 02139}
\author{David Woolf}
\affiliation{School of Engineering and Applied Sciences,
Harvard University, Cambridge, MA 02139}
\author{Pui-Chuen Hui}
\affiliation{School of Engineering and Applied Sciences,
Harvard University, Cambridge, MA 02139}
\author{Eiji Iwase}
\affiliation{School of Engineering and Applied Sciences,
Harvard University, Cambridge, MA 02139}
\author{Alexander P. McCauley}
\affiliation{Department of Physics,
Massachusetts Institute of Technology, Cambridge, MA 02139}
\author{Federico Capasso}
\affiliation{School of Engineering and Applied Sciences,
Harvard University, Cambridge, MA 02139}
\author{Marko Loncar}
\affiliation{School of Engineering and Applied Sciences,
Harvard University, Cambridge, MA 02139}
\author{Steven G. Johnson}
\affiliation{Department of Mathematics,
Massachusetts Institute of Technology, Cambridge, MA 02139}

\begin{abstract}  
  We propose an optomechanical structure consisting of a
  photonic-crystal (holey) membrane suspended above a layered
  silicon-on-insulator substrate in which resonant bonding/antibonding
  optical forces created by externally incident light from above
  enable all-optical control and actuation of stiction effects induced
  by the Casimir force. In this way, one can control how the Casimir
  force is expressed in the mechanical dynamics of the membrane, not
  by changing the Casimir force directly but by optically modifying
  the geometry and counteracting the mechanical spring constant to
  bring the system in or out of regimes where Casimir physics
  dominate. The same optical response (reflection spectrum) of the
  membrane to the incident light can be exploited to accurately
  measure the effects of the Casimir force on the equilibrium
  separation of the membrane.
\end{abstract}

%\ocis{(190.2620) Nonlinear optics: frequency conversion; (230.4320) Nonlinear optical devices}

\maketitle

Casimir forces between neutral objects arise due to quantum and
thermal fluctuations of the electromagnetic field and, being
ordinarily attractive, can contribute to the failure (stiction) of
micro-electromechanical systems
(MEMS)~\cite{serry98,Buks01:mems,Capasso07:review}. In this letter, we
propose a scheme for controlling and measuring the Casimir force
between a photonic-crystal membrane and a layered substrate that
exploits the resonant optomechanical forces created by evanescent
fields~\cite{Povinelli05,Kippenberg08,Thourhout10} in response to
external, normally incident light~\cite{RodriguezMc11OE}. Our
numerical experiments reveal a sensitive relationship between the
equilibrium separation of the membrane and the Casimir force, as well
as demonstrate low-power optical control over stiction
effects. (Although our focus is on stiction induced by the Casimir
force, similar results should also apply in circumstances involving
electrostatic forces)~\cite{Capasso07:review}.  Casimir forces have
most commonly been measured in cantilever experiments involving
sphere--plate geometries~\cite{milton04,Lamoreaux05,Capasso07:review},
with some exceptions~\cite{RodriguezCa11}, in which the force is often
determined by measuring its gradient as a function of object
separation. Another approach involves measuring the dynamic response
of a plate to mechanical modulations induced by an electrostatic
voltage~\cite{hochan2,Capasso07:review}. Here, we consider an
alternative scheme in which the Casimir force is determined instead
via \emph{pump}--\emph{probe} measurements of the optically tunable
equilibrium separation between a membrane and a substrate, a proof of
concept of an approach to all-optical control and actuation of
opto-micromechanical devices susceptible to
stiction~\cite{Buks01:mems,Capasso07:review,Manipatruni09,Pernice10}.

\begin{figure}[t!]
\centering
\includegraphics[width=1.0\columnwidth]{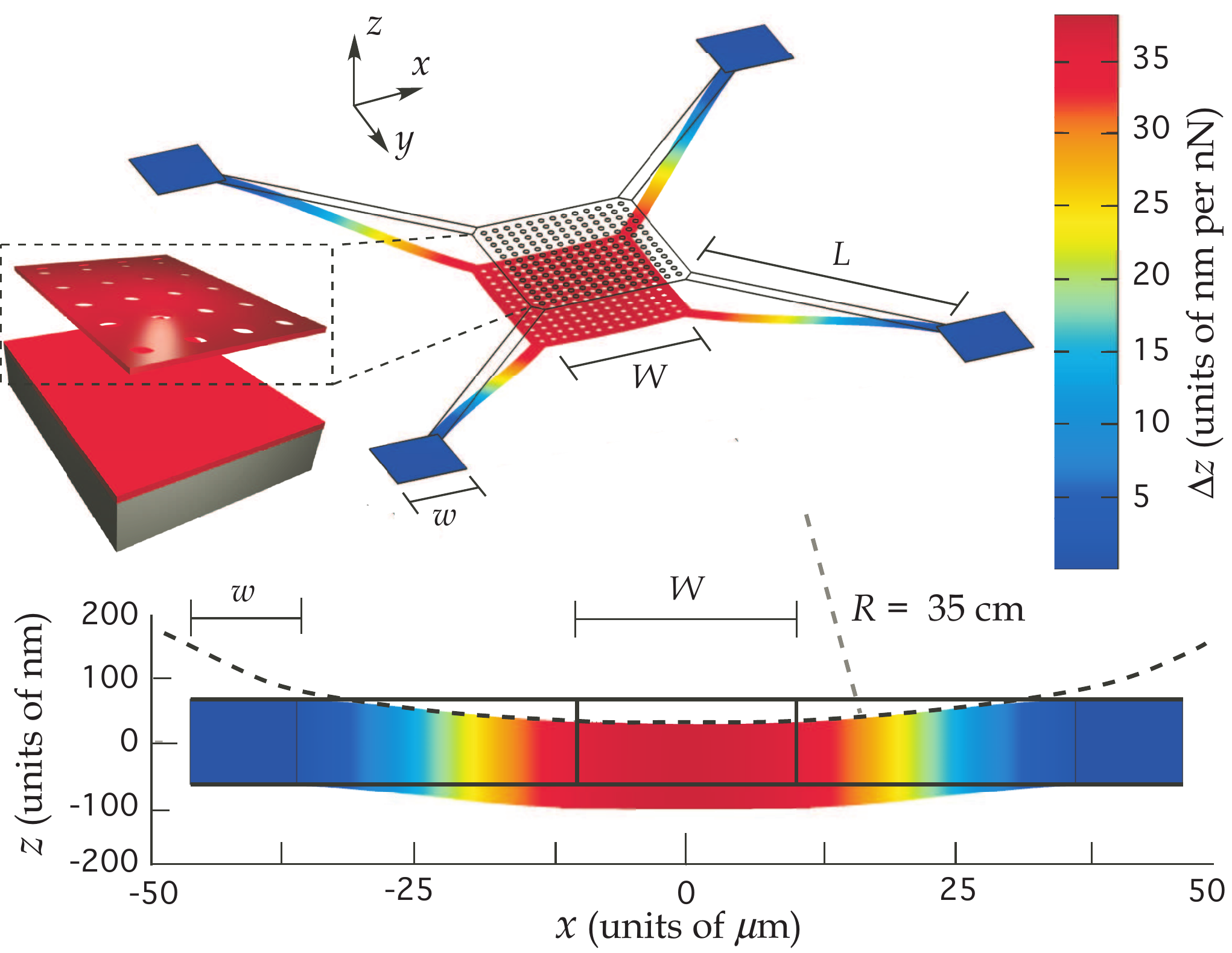}
\caption{Schematic of single-membrane structure (thickness $h =
  130$~nm and size $W = 23.4~\mu$m), designed so that normally
  incident light from above ($+z$ direction) induces resonant optical
  forces on the membrane. Also shown is a colorbar of the membrane
  displacement due to an impinging 1~nN force. Notice that the arms
  supporting the membrane (length $L = 35~\mu$m) bend significantly
  more than the membrane: the total bending of the membrane is
  $\approx 38$~nm, while the membrane's center--corner height
  difference is a mere $2.7$~nm, corresponding to an effective radius
  of curvature $R \approx 35$~cm.}
\label{fig:fig1}
\end{figure}

We focus on the optomechanical structure shown in \figref{fig1}, and
which we first examined in~\citeasnoun{RodriguezMc11OE}: a silicon
membrane of thickness $h = 130$~nm and width $W =
23.4~{\mu}\mathrm{m}$ perforated with a square lattice of air holes of
diameter $260$~nm and period $650$~nm, suspended above a layered
substrate---a silicon film of thickness $h = 130$~nm on a silica (SOI)
substrate---by four rectangular supports of length $L = 35~{\mu}$m and
cross-sectional area $130~\mathrm{nm} \times 2~\mu\mathrm{m}$.

In what follows, we consider quasistatic membrane deformations induced
by static (and spatially uniform) optical/Casimir forces, so it
suffices to study the fundamental mechanical mode and corresponding
frequency $\Omega_m$ of the membrane. The mode profile is illustrated
in~\figref{fig1} and consists of an approximately flat membrane with
deformed supports, making this structure less susceptible to optical
losses stemming from curvature [\figref{fig1} caption]. For the
particular configuration studied here, we found $\Omega_m \approx
63$~kHz, corresponding to a mechanical spring constant $\kappa_m
\approx 5 \times 10^{-2}$~N/m.

\begin{figure}[t!]
\centering
\includegraphics[width=1.0\columnwidth]{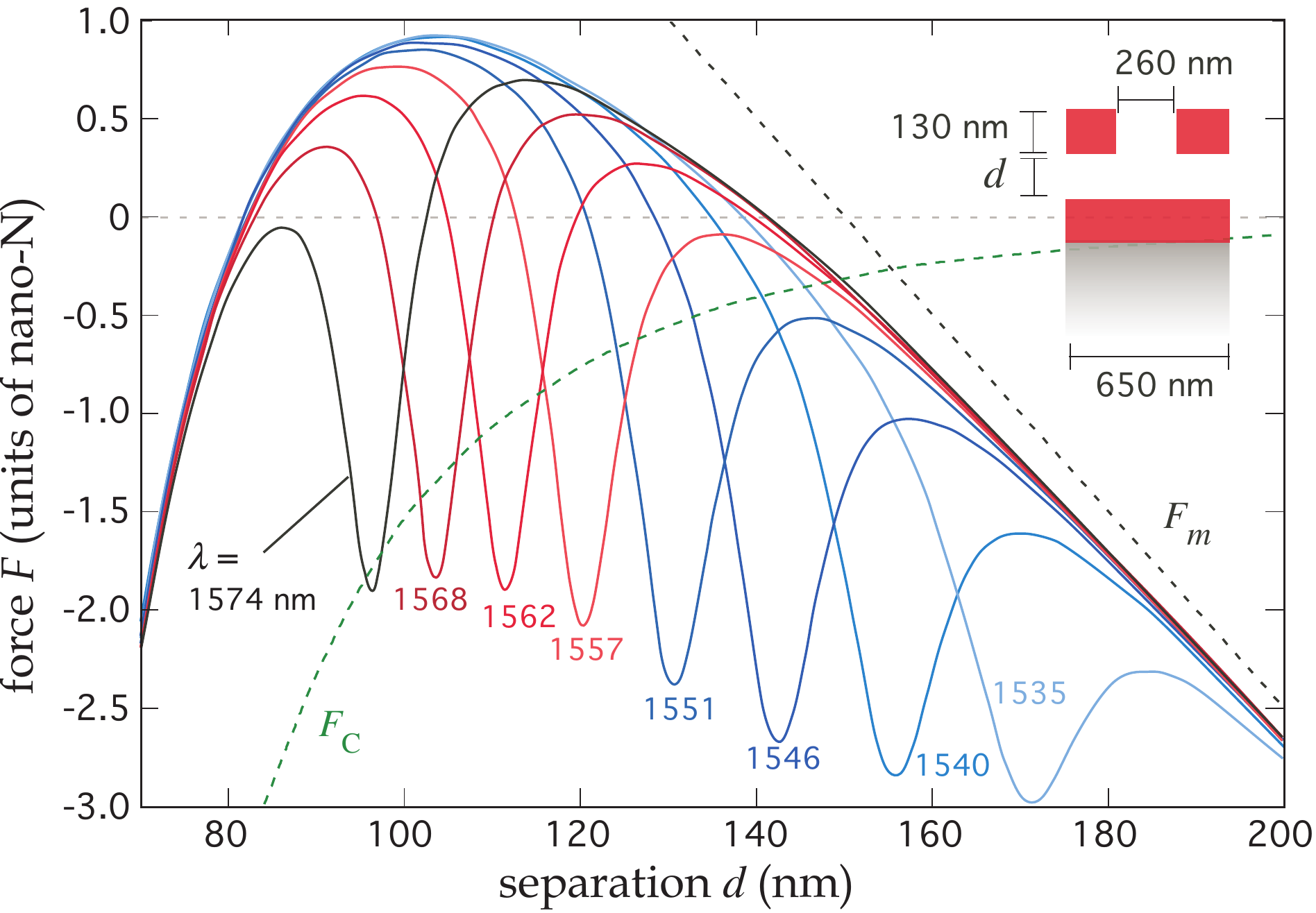}
\caption{Total force $F = F_m + F_c + F_o$ on the membrane
  of~\figref{fig1} (cross-section shown on the inset), initially
  suspended at $d_0 = 150$~nm, as a function of membrane separation
  $d$. $F_m$ is the mechanical restoring force (dashed black), $F_c$
  the Casimir force (dashed green) and $F_o$ the optical force induced
  by normally-incident light of power $P=10~\mathrm{m}W$ and
  wavelength $\lambda$. $F$ is plotted for different $\lambda$.}
\label{fig:fig2}
\end{figure}

Away from a desired initial mechanical membrane--surface separation
$d_0$, here chosen to be $d_0 = 150$~nm, and in the absence of optical
forces, the membrane will experience two forces, plotted
in~\figref{fig2} as a function of separation $d$: a restoring
mechanical force $F_m = \kappa_m (d_0-d)$ that increases linearly with
separation $d$ (dashed black line), and the attractive,
monotonically-decaying Casimir force $F_c$ (dashed green line). $F_c$
was computed via the standard proximity-force
approximation~\cite{milton04,Genet08}, which we have checked against
exact time-domain
calculations~\cite{RodriguezMc09:PRA,McCauleyRo10:PRA} and found to be
accurate to within $3\%$. $F_c$ has two major effects on the membrane:
First, it leads to a new equilibrium separation $d_c \approx 140$~nm;
Second, it creates an unstable equilibrium at a smaller separation
$d_u \approx 80$~nm, determined by the competition between $F_m$ and
$F_c$, below which the membrane will stick to the substrate.  We
propose that the Casimir force can be measured by optically
controlling the equilibrium separation in real time by illuminating
the membrane with normally incident light at a tunable wavelength
$\lambda$, which creates a resonant force and allows one to
dynamically determine the Casimir-induced threshold for
stiction~\cite{RodriguezMc11OE}. In particular, \figref{fig2} also
plots the total force on the membrane $F = F_m + F_c + F_o$, where
$F_o$ is the single-$\lambda$ optical force on the membrane induced by
incident light of power $P = 10~\mathrm{m}W$. Here, the strong
$\lambda$-dependence of $F_o$ is exploited to obtain large and tunable
attractive (bonding) forces at any desired $d$ (solid lines), allowing
us to control the equilibrium separation of the
membrane~\cite{RodriguezMc11OE}. As shown, slowly increasing $\lambda$
from $\lambda \approx 1520$~nm to $\lambda \approx 1581$~nm causes
$d_c$ to decrease and come arbitrarily close to $d_u$.

\begin{figure}[t!]
\centering \includegraphics[width=1.0\columnwidth]{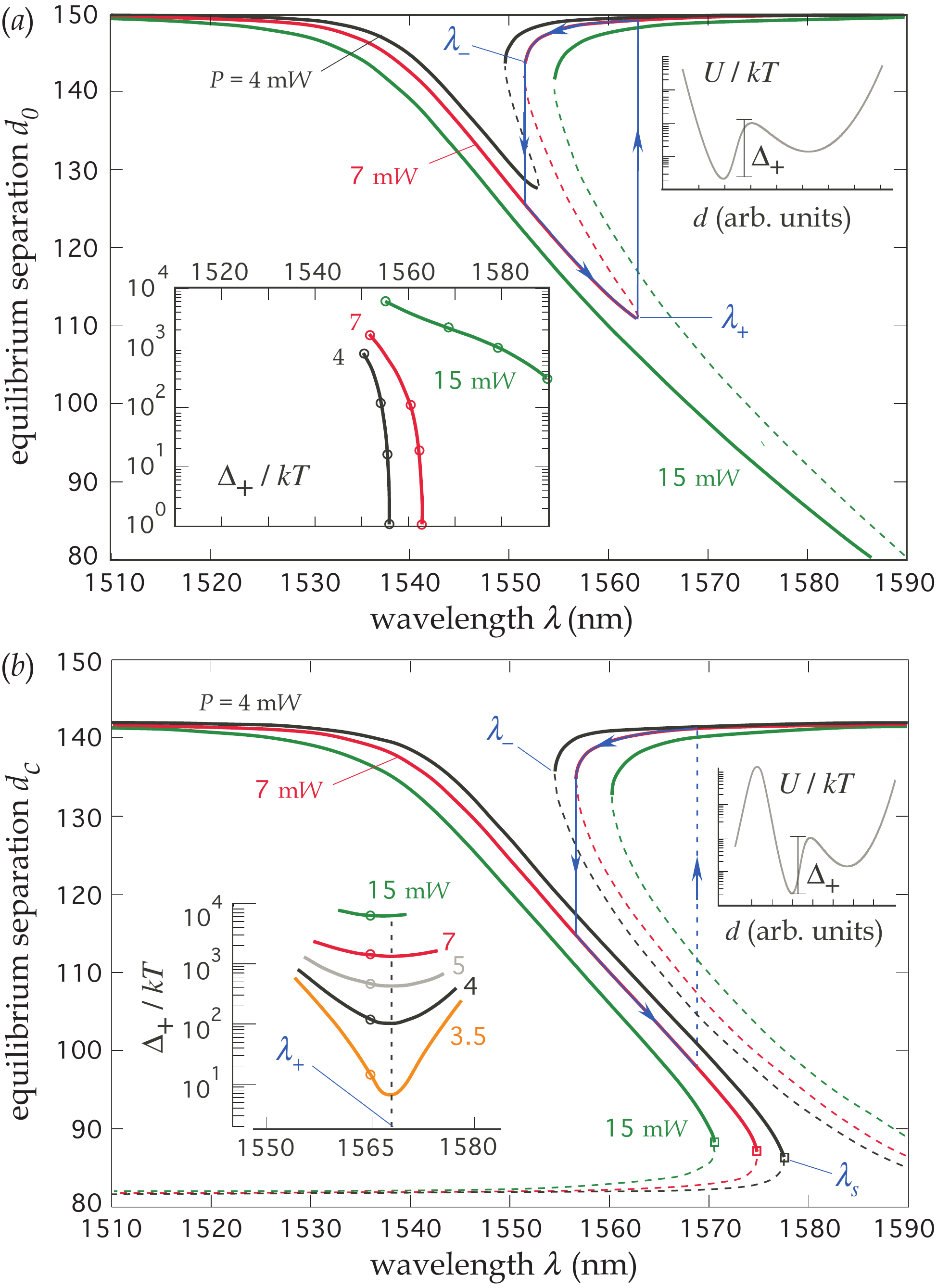}
\caption{Stable (solid) and unstable (dashed) membrane equilibria
  separations in response to normally-incident light of power $P$, as
  a function of optical wavelength $\lambda$, plotted for multiple
  $P$, excluding (a) and including (b) the Casimir force $F_c$. The
  right insets are schematic illustrations of the energy landscape as
  a function of membrane separation $d$, indicating the energy barrier
  $\Delta_+ / kT$ separating the lower and higher stable
  equilibria. The left insets plot $\Delta_+ / kT$ as a function of
  $\lambda$ plotted for multiple $P$.}
\label{fig:fig3}
\end{figure}

\Figref{fig3} quantifies the effect of $F_o$ and $F_c$ on the
equilibrium of the membrane. In particular, \figref{fig3}(a) and
\figref{fig3}(b) plot the equilbrium separation, $d_c$ and $d_o$, in
the absence ($F_c = 0$) and presence ($F_c \neq 0$) of the Casimir
force, respectively, as a function of $\lambda \in [1510, 1590]$~nm,
for light incident at various $P$. When $F_c = 0$ [\figref{fig3}(a)],
increasing $\lambda$ has two main effects on the membrane: First, the
equilibrium separation decreases; Second, two bifurcation wavelengths,
denoted by $\lambda_{\pm}$ (indicated in the figure), are created due
to the presence of two additional (stable and unstable) equilibria,
leading to bistability and hysteresis effects as $\lambda$ is
varied~\cite{Dorsel83,Buks01:mems}. This is illustrated by the blue
curve in the particular case of $P=7~\mathrm{m}W$: if $\lambda$ is
slowly increased from $\lambda \approx 1510$~nm, $d_o$ decreases until
$\lambda \to \lambda_+ \approx 1562$~nm, at which point the stable and
unstable equilibria bifurcate and the position of the membrane changes
dramatically from $d_o = 112$~nm to $d_o \approx 150$~nm. Decreasing
$\lambda$ below $\lambda_+$ after the transition causes the membrane
to tranverse a different path, leading to another dramatic change in
$d_o$ as $\lambda \to \lambda_- \approx 1551$~nm from above. The same
is true at other $P$, although of course the corresponding
$\lambda_{\pm}$ will change. The presence of $F_c > 0$
[\figref{fig3}(b)] affects the membrane's response to $F_o$ in
important ways: For small $P$, $F_o$ is too weak and therefore $d_c$
is too large for $F_c$ to have a strong effect on the membrane (the
membrane reaches $\lambda_+$ before it can feel the Casimir force). At
larger $P > P_c \approx 3~\mathrm{m}W$, however, $d_c$ and
$\lambda_{-}$ are greatly affected by $F_c$ and there is no longer any
optical bistability: the bifurcation point $\lambda_+$ is instead
replaced with a new bifurcation point $\lambda_s > \lambda_+$, arising
from the lower (stable) and Casimir-induced (unstable) equilibria,
leading to stiction rather than a jump in the membrane separation as
$\lambda \to \lambda_s$.

In an experiment, the presence of Brownian motion will cause membrane
fluctuations about the stable equilibria, and these can lead to
dramatic transitions in the position of the membrane, from one stable
equilibrium to another, and even to stiction, as the membrane moves
past the energy barrier $\Delta$ separating the various stable
equilibria; the right insets of \figref{fig3} illustrate the energy
landscape of the membrane. The average lifetime $\tau$ of a metastable
equilibrium is proportional to $\exp(\Delta/kT)$, which explains why
they are rarely observed in stiff mechanical systems where $\Delta /
kT \gg 1$,~\cite{Buks01:mems} but in our case $\tau$ can be made
arbitrarily small by exploiting $F_o$: the barrier $\Delta_{\pm}$
between the two stable equilibria $\to 0$ as $\lambda \to
\lambda_{\pm}$, as shown by the left inset of \figref{fig3}(a).  The
presence of $F_c$ has a dramatic effect on these thermally induced
transitions. In particular, even though there is no bifurcation point
$\lambda_+$, the barrier from the smallest to larger stable $d_c$
(denoted by $\Delta_+$) can be made arbitrarily small and varies
non-monotonically with $\lambda$: as $\lambda$ is increased from
$\lambda_- \to \lambda_s$, $\Delta_+$ decreases and then increases as
$\lambda$ passes through a critical $\lambda_+$ [indicated
  in~\figref{fig3}(b)]. This potential ``dip'' gets deeper as $P$
decreases [as shown in the left inset of \figref{fig3}(b)], making it
easier for the membrane to transition to the larger $d_c$. Thus, for
sufficiently small $P$, the \emph{rate} at which $\lambda$ is
increased determines whether the membrane transitions upwards near
$\lambda_+$ or sticks to the substrate near $\lambda_s$: if
$\mathrm{d}\lambda/\mathrm{d}t \gg \lambda/\tau$ near $\lambda_+$, the
upward transition is ``frustrated''. This creates a hysteresis effect
[illustrated by the blue curve in~\figref{fig3}(b)] where the upward
transition (dashed blue line) can occur only due to thermal
fluctuations and whether or not this occurs will depend on $P$ and
$\mathrm{d}\lambda/\mathrm{d}t$. For $P$ smaller than a critical $P_c
\approx 3~\mathrm{m}W$, the lower stable and higher unstable $d_c$
merge, leading to two additional bifurcations (not shown), and it
becomes impossible to continuously change $\lambda$ to obtain a
transition from stable suspension into stiction [instead, the optical
  bistability behavior of \figref{fig3}(a) is observed]. This ability
to tune the stiction barrier through $F_o$ remains an unexplored
avenue for experimentally gauging the impact of Casimir and other
stiction forces on the operation of optomechanical systems.

\begin{figure}[t!]
\centering
\includegraphics[width=0.94\columnwidth]{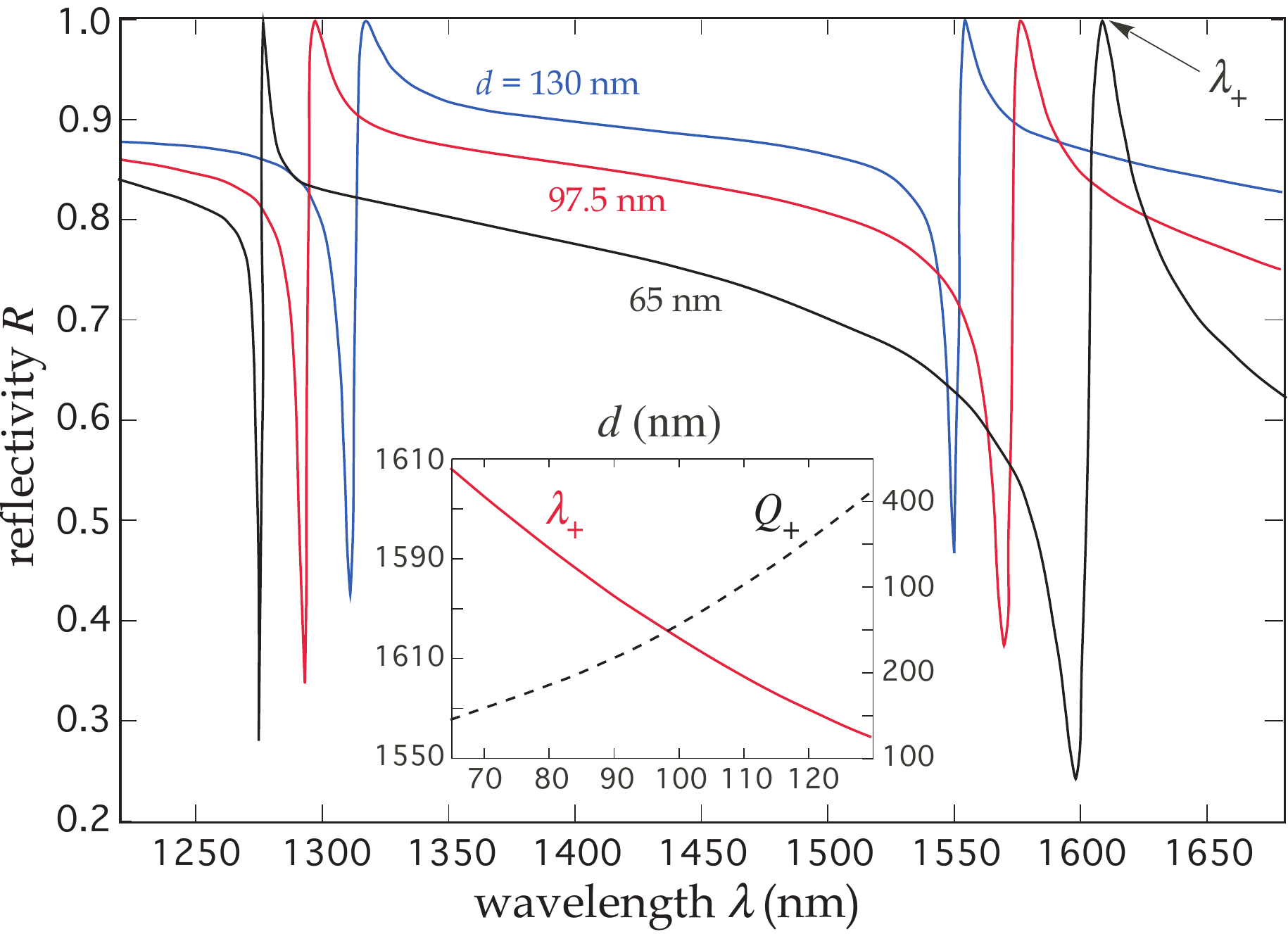}
\caption{Reflectivity $R$ as a function of optical wavelength
  $\lambda$ at various membrane separations $d$. The inset shows the
  peak wavelength $\lambda_+$ (solid red) and lifetime $Q_+$ (dashed
  black) as a function of $d$.}
\label{fig:fig4}
\end{figure}

Our results thus far demonstrate a sensitive dependence of $d_c$ on
$F_c$ and $\lambda$. However, determining $F_c$ accurately rests on
the ability to determine $d_c$ accurately, and we propose to measure
the latter interferometrically via the (broadband) optical response of
the membrane~\cite{RodriguezMc11OE}. \Figref{fig4} plots the
reflectivity $R$ of the membrane as a function of $\lambda$ at various
$d$, showing the presence of multiple reflection peaks at positions
$\lambda_{\pm}$ (indicated in the figure) that shift as $d$ is
varied---these correspond to the bonding ($+$) and antibonding ($-$)
resonances that allowed us earlier to control the membrane's
equilibrium separation. The inset plots the $\lambda_+$ and
corresponding lifetime $Q_+$ of the bonding mode as $d$ is varied,
revealing a large change in $\mathrm{d}\lambda_+/\mathrm{d}d \in [0.5,
  1.2]$ and $\mathrm{d}Q_+/\mathrm{d}d \in [5,12]$~nm$^{-1}$ over the
entire $d$-range.

The basic phenomena described here are by no means limited to the
particular realization of this geometry, nor to our choice of initial
equilibrium position, and we believe that similar and more pronounced
effects should be present in other configurations. For example,
dramatically lower $P$ can be obtained by increasing the $Q$ of the
membrane resonances, beyond the mere $Q \sim 10^2$ here, e.g. by
decreasing the radii of the air holes (limited by losses). We note
that antibonding (repulsive) forces can also be exploited, in
conjuncion with bonding forces, to overcome prohibitive stiction
effects in similar optomechanical systems~\cite{MaboudianHowe97},
e.g. as an antistiction feedback mechanism~\cite{RodriguezMc11OE}.

This work was supported by the Defense Advanced Research Projects
Agency (DARPA) under contract N66001-09-1-2070-DOD.

%\bibliographystyle{optlett}
%\bibliography{photon}

\end{document}